\documentclass[prb,twocolumn,superscriptaddress,preprintnumbers,amsmath,amssymb]{revtex4}
\usepackage[dvips]{graphicx}% Include  files
\usepackage{dcolumn}% Align table columns on decimal point
\usepackage{color}
\usepackage{bm}% bold math

\newcommand{\be}{\begin{equation}}
\newcommand{\la}{\langle}
\newcommand{\ra}{\rangle}
\newcommand{\ee}{\end{equation}}
\newcommand{\bea}{\begin{eqnarray}}
\newcommand{\eea}{\end{eqnarray}}
\newcommand{\bd}{\begin{displaymath}}
\newcommand{\ed}{\end{displaymath}}

\renewcommand{\r}{\rho}
\newcommand{\s}{\sigma}

\newcommand{\vf}{\varphi}

\renewcommand{\o}{\omega}

\newcommand{\lf}{\left}
\newcommand{\rg}{\right}

\newcommand{\TKKcomp}{Department of Applied Physics/COMP, Aalto University, P.O.~Box 14100, FI-00076 AALTO, Finland}
\newcommand{\TKKltt}{Low Temperature Laboratory, Aalto University, 
  P.O.~Box 13500, FI-00076 AALTO, Finland}

\begin{document}
\title{Decoherence of adiabatically steered quantum systems}

\author{P. Solinas}
\affiliation{\TKKcomp} 

\author{M. M\"ott\"onen}
\affiliation{\TKKcomp}
\affiliation{\TKKltt}
\affiliation{Australian Research Council Centre of Excellence for Quantum Computer
    Technology, School of Electrical Engineering \& Telecommunications,
    University of New South Wales, Sydney, New South Wales 2052, Australia}
    
\author{J. Salmilehto}
\affiliation{\TKKcomp}

\author{J. P. Pekola}
\affiliation{\TKKltt}

\begin{abstract}
We study the effect of Markovian environmental noise on the dynamics of a two-level quantum system which is steered adiabatically by an external driving field.
We express the master equation taking consistently into account all the contributions to the lowest non-vanishing order in the coupling to the Markovian environment.
We study the master equation numerically and analytically and we find that, in the adiabatic limit, a zero-temperature environment does not affect the ground state evolution. 
As a physical application, we discuss extensively  how the environment affects Cooper pair pumping.
The adiabatic ground state pumping appears to be robust against environmental noise.
In fact, the relaxation due to the environment is required to avoid the accumulation of small errors from each pumping cycle.
We show that neglecting the non-secular terms in the master equation leads to unphysical results, such as charge non-conservation.
We discuss also a possible way to control the environmental noise in a realistic physical setup and its influence on the pumping process.
\end{abstract}

\maketitle 

\section{ Introduction}
During the past three decades along with a greatly increased interest in quantum mechanics and its applications, 
we have witnessed a growing demand for accurate control of quantum systems.
However, these systems are difficult to manipulate and very sensitive to external noise.

A possible solution to the accurate control problem may be the adiabatic manipulation.
This approach not only eliminates the need for fine tuning the time control, but 
recently it has been proposed to solve complex computational problems \cite{farhi01} and  shown to be able 
to produce any unitary transformation. \cite{ad_vs_dyn}
Within the same framework, it is also possible to perform non-trivial periodic control to manipulate 
the quantum system by geometric means.
This approach originates from the famous work of Berry \cite{berry84} but later it has been extended to more general cases
\cite{wilczek84} and analyzed for its possible applications in the so-called geometric quantum computing. \cite{falci00,zanardi99,unanyan99,duan01}
The main advantage is that the geometric dependence of the applied unitary operator renders  it robust against fast fluctuations in the control signal.
\cite{dechiara03,solinas04}
The main drawback of this proposal is that to exploit all the advantages, the evolution time has to be long with respect 
to the time scale of the system.
However, the evolution time cannot be increased indefinitely because the system undergoes 
dephasing and relaxation leading potentially to serious problems in maintaining the quantum state.
Even if some efforts have been made to clarify the effect of the environment and its possible elimination
in the context of geometric quantum computation, \cite{parodi07,florio06} a complete analysis is still missing.

In this paper, we consider a two-level system initially in the ground state and study how it is influenced by the environment
and an adiabatically changing external driving field.
The time-dependent control makes the usual treatment \cite{cohen-tannoudji} by means of Markovian master equation non-trivial. \cite{pekola09} 
Similar problems have been studied with different approaches \cite{calarco03,roszak05,caillet07} in the aim to estimate the decoherence effect induced by the environment.
Here, we use a particular master equation derived taking consistently into account the drive, the environment and the combined effect.
In particular, we include the usually neglected secular terms and find that, in our case, they be handled with extreme care since it can lead to unphysical results.
Our result that zero-temperature environment does not affect the ground state adiabatic dynamics suggests that 
geometric quantum computation in the ground state \cite{pirkkalainen10} could be robust against decoherence.
The analytical derivation of the solutions of the master equation for different coupling and adiabatic evolution is discussed in detail. Similar results are obtained in the more realistic case of finite temperature environment. 

As a physical example, we study Cooper pair pumping \cite{niskanen03} and discuss in detail the effect of the environment in the pumping process.
In this case, we find that the physical observables are robust during the ground state pumping.  
However, neglecting the non-secular terms in the master equation leads to a decrease in the predicted pumped charge and a difference in the average charge pumped through the two junctions, i.e., charge non-conservation. 
These results are obtained analytically by direct calculation and confirmed by numerical analysis. %% here
We propose a way to implement {\it in situ} the environmental noise by coupling a thermal resistor to the system through an array of SQUIDs. The main advantage of this proposal is that, by controlling the flux through the SQUID array, we can experimentally change the effective noise spectrum of the environment.

This paper is organized as follows. In Sec. \ref{sec:model} the model is described.
In Sec. \ref{sec:MEQ}, we write the master equation, solve it in the quasi-stationary limit, and discuss different approximations.
In Sec. \ref{sec:physical_model}, we apply the present model to Cooper pair pumping in the presence of environment and discuss different 
regimes, the effect of temperature, and the implication of the secular approximation.
Furthermore, we describe a way to engineer the environment {\it in situ}, and calculate the pumped charge.
Section \ref{sec:conclusions}  concludes the paper.

\section{Model}
\label{sec:model}

We consider a quantum system  subject to external time-dependent control fields and interacting with the environment.
The total Hamiltonian is 
$H(t)=H_S(t)+H_E+ V$, where $H_S(t)$ denotes the time-dependent system Hamiltonian with non-degenerate eigenvalues, 
$H_E$ is the environment Hamiltonian  and $V$ is the system-environment coupling. 
We assume that $V=Y \otimes X$, where $X$ is the environment and $Y$ is the system part of the coupling operator.
The simplest way to study the dynamics of the system is to use the instantaneous eigenbasis of $H_S(t)$, i.e., the \emph{adiabatic basis} composed of states  $|\psi_n(t) \rangle $ with energies $E_n$ such that $H_S(t)\lf|\psi_n(t)\rg>=E_{n}(t)\lf|\psi_n(t)\rg>$.
We denote by $D(t)$ the transformation from a given fixed basis to the adiabatic one.
To properly take into account the time-dependence of the basis, we study the dynamics of $\tilde{\rho}_{\rm tot}=D^\dagger(t)\rho_{\rm tot} (t) D(t)$ in the fixed basis. Here $\rho_{\rm tot}(t)$ is the usual density operator of the total system in the Schr\"odinger picture.
The evolution of $\tilde{\rho}_{\rm tot}$ is governed by the effective Hamiltonian
\begin{equation}
\tilde H^{(1)}(t)=\tilde H_S(t)+ \hbar w(t) + \tilde V(t)+ H_E , 
\label{eq:Ham1}
\end{equation}
where $\tilde H_S(t)= D^\dag(t)H_S(t)D(t)$, $\tilde{V}(t)=D^\dag(t) V D(t)=\tilde Y(t) \otimes X$, and $w = -iD^\dagger(t) \dot{D}(t)$.

We define the local adiabatic parameter as $\alpha(t)=\hbar ||w(t)||/| \Delta(t)|$ where $\Delta(t)$ is the instantaneous minimum energy gap in the spectrum and  $||w(t)||= {\rm Tr}_S \sqrt{w^\dagger(t) w(t)}$ is the trace norm of $w(t)$.  
This instantaneous adiabatic parameter gives an accurate estimate of the local degree of adiabaticity of the evolution.
Denoting with $T_p$ the time interval for the adiabatic evolution and setting $\Delta_{min} = {\rm min} [\Delta(t)]$,
a rough estimate of $\alpha$ is given by $\bar{\alpha}= \hbar / (\Delta_{min} T_p)$ which represents a global adiabatic parameter.
The $w(t)$ scales as $1/T_p$ and then, for adiabatic evolution ($\alpha \ll 1 $) it is usually neglected.
\cite{sarandy05} However, 
as discussed in Ref. \cite{pekola09}, it can be crucial to include its influence for the correct description of the dynamics of the system.
In fact, the full Hamiltonian (\ref{eq:Ham1}) suggests that the environment leads to relaxation to the instantaneous eigenstates of  $\tilde H_S(t)+ \hbar w(t)$.
If we neglect $w(t)$, the environment produces relaxation in the adiabatic basis of $\tilde H_S(t)$ which usually differs from the 
correct one at the order $O(\alpha)$. As we will discuss in detail, this contribution can be important in determining the expectation value
of a physical observable.

\section{ Master Equation}
\label{sec:MEQ}

Let us study the dynamics of a generic two-level system steered by a time-dependent Hamiltonian and 
subject to environmental noise.
If the evolution is adiabatic and the system interacts weakly with the environment, we can treat $\hbar w(t) + \tilde V(t)$ 
as a perturbation and derive a master equation for the density matrix in the interaction picture. \cite{pekola09}
Up to the first order in $\alpha$ and the second in the system-environment coupling, we obtain
\begin{widetext}
\begin{eqnarray} \label{me1}
&&\frac{d\tilde{\rho}_I(t)}{dt} = i [\tilde{\rho}_I(t),w_I(t)] 
-\frac{1}{\hbar^2}{\rm Tr}_E \lf\{ \int_0^t dt' \big[[\tilde{\rho}_I(t)\otimes\r_E,\tilde{V}_I(t')],\tilde{V}_I(t)\big] \rg\} + \nonumber\\
&& +\frac{i}{\hbar^2}{\rm Tr}_E \lf\{ 
\int_0^t dt' \int_{0}^{t'}dt''\Big[~\big[\tilde{\rho}_I(t)\otimes\r_E,[w_I(t'),\tilde{V}_I(t'')]~\big],\tilde{V}_I(t)\Big] \rg\} ,
\label{eq:formal_MEQ}
\end{eqnarray}
\end{widetext}
where  $\tilde{\rho}_I(t)$ is the density matrix of the system in the interaction picture, ${\rm Tr}_E$  indicates trace over the
environmental degrees of freedom and ${\rho}_E$ is the stationary density operator of the environment. The interaction picture operators are defined as
$\tilde O_I(t)=e^{i H_E t/\hbar} U^\dag_S(t,0)\tilde O (t)U_S(t,0)e^{-i H_E t/\hbar}$,
where $U_S(t,0)=\mathcal{T} e^{-i/\hbar \int_0^{t}\tilde{H}_S(\tau)d\tau}$ is the time-evolution operator, with $\mathcal{T}$ denoting time-ordering.
In this frame, we can address and interpret the contributions of Eq. (\ref{eq:formal_MEQ}).
The first contribution on the right is the driving term without system-environment interaction, the second one leads to the standard dissipative contribution of the Bloch-Redfield theory.
The third contribution is a cross-term of the drive and dissipation. 
This last term is usually  neglected as discussed above. \cite{whitney05}

Denoting the ground and excited state of $H_S$ as $|g\rangle$ and $|e\rangle$, respectively, the master equation arising from Eq. (\ref{eq:formal_MEQ}) in the Schr\"odinger picture assumes the form
\begin{widetext}
\begin{eqnarray}
 \dot{\rho}_{gg} &=& -2 \Im \mbox{m} (w_{ge}^* \rho_{ge})+
 S(\omega_0 ) \left|m_2\right|{}^2-[S(-\omega_0 )+S(\omega_0 )]\left |m_2\right|{}^2 \rho _{gg} +2 \left[\Im \mbox{m}\left(m_2\right)
   \Im \mbox{m}\left(\rho _{ge}\right)+\Re \mbox{e}\left(m_2\right) \Re \mbox{e}\left(\rho _{ge}\right)\right] S(0) m_1 \nonumber\\ &&
   -2 \frac{2 S(0)-S(-\omega_0 )-S(\omega_0 )}{ \omega_0 }\left\{ [\Im \mbox{m}\left(m_2\right) \Im \mbox{m}\left(w_{ge}\right)+\Re \mbox{e}\left(m_2\right) \Re \mbox{e}\left(w_{ge}\right)\right]
   \left[\Im \mbox{m}\left(m_2\right) \Im \mbox{m}\left(\rho _{ge}\right)+\Re \mbox{e}\left(m_2\right) \Re \mbox{e}\left(\rho _{ge}\right)\right] \} \nonumber\\ &&
   +2 \frac{2 S(0)-S(-\omega_0 )-S(\omega_0 )}{\omega_0 }m_1 \left\{\Im \mbox{m}\left(m_2\right) \Im \mbox{m}\left(w_{ge}\right)+\Re \mbox{e}\left(m_2\right)
   \Re \mbox{e}\left(w_{ge}\right)\right\} \rho _{gg}  \nonumber\\§ &&
   -2\frac{S(0)-S(\omega_0 ) }{\omega_0 }m_1\left\{\Im \mbox{m}\left(m_2\right)
   \Im \mbox{m}\left(w_{ge}\right)+\Re \mbox{e} \left(m_2\right) \Re \mbox{e}\left(w_{ge}\right) \right\},
  \label{eq:Eq1}
\end{eqnarray}
\end{widetext}
and
\newpage
\begin{widetext}
\begin{eqnarray}
 \dot{\rho}_{ge} &=& i w_{ge} (2 \rho_{gg}-1) + i (w_{ee}-w_{gg}) \rho_{ge}+i \omega_0 \rho_{ge}
 -S(\omega_0 ) m_1 m_2 +[S(-\omega_0 ) +S(\omega_0 ) ] m_1 m_2 \rho _{gg} -2 S(0)  m_1^2 \rho _{ge} \nonumber\\
  &&-i [S(-\omega_0 )+S(\omega_0 )] m_2 [ \Re\mbox{e}(m_2) \Im \mbox{m}(\rho _{ge})- \Im \mbox{m}(m_2) \Re\mbox{e}(\rho_{ge})] \nonumber\\ 
 &&- 2 \frac{2 S(0)-S(-\omega_0 )-S(\omega_0 )}{\omega_0 }  m_1^2 w_{ge} \rho _{gg} \nonumber\\
 &&+ 2 \frac{S(0)-S(\omega_0 )}{\omega_0} m_1^2 w_{ge}- i m_2 \frac{S(-\omega_0 )-S(\omega_0 )}{\omega_0} \{ \Im \mbox{m}(m_2)
   \Re \mbox{e}(w_{ge})- \Re \mbox{e}(m_2)  \Im \mbox{m}(w_{ge}) \} \nonumber\\
 & &-2\frac{ 2 S(0)-S(-\omega_0 )-S(\omega_0 )}{\omega_0 }  m_1 \{ i m_2 [\Im \mbox{m}(w_{ge}) \Re \mbox{e}(\rho_{ge})-\Re \mbox{e}(w_{ge}) \Im \mbox{m}(\rho _{ge}))] \nonumber\\
 &&-[\Im \mbox{m}\left(m_2\right)   \Im \mbox{m}(w_{ge})+\Re \mbox{e}(m_2)s \Re \mbox{e}(w_{ge})] \rho_{ge} \} ,
\label{eq:Eq2}
\end{eqnarray}
\end{widetext}
where $\hbar \omega_0$ is the (instantaneous) energy gap between $|g\rangle$ and $|e\rangle$, $m_1=Y_{gg}(t)/\hbar=-Y_{ee}(t)/\hbar$, $m_2=Y_{ge}(t)/\hbar$, and $S(\pm \omega_0) $ and $S(0)$ are obtained from the power spectrum of the noise $S(\omega)=\int_{-\infty}^\infty\la X_{I}(\tau)X_I(0)\ra e^{i\o\tau}d\tau$. \cite{note_m}
Here, the matrix element of a generic operator $O$ is denoted by $O_{kl}=\langle k|O|l\rangle$ (with $k,l=g,e$) except for $w_{kl}=-i \langle k|\dot{l}\rangle$. 
Notice that the present equations generalize the ones presented in Ref. \cite{pekola09} because they are obtained for a generic system-environment coupling operator.

In the derivation of equations (\ref{eq:Eq1}) and (\ref{eq:Eq2}), we have implicitly assumed that our system is in the Markovian regime.
This implies that the bath correlation time $\tau$ is short compared to the relaxation time of the system. 
In addition, since the evolution is adiabatic, $\omega_0$, $m_1$, $m_2$, and $w$ change slowly in time and we can approximate $w(t+ \tau ) \approx w(t)$.

In writing the above equation we have neglected the corrections due to the Lamb shift. However, we expect it to have small influence on the dynamics of the system in the adiabatic limit. A detailed analysis is under way. \cite{lamb_shift_note}

Notice that Eqs.~(\ref{eq:Eq1}) and~(\ref{eq:Eq2}) are, strictly speaking, valid only for adiabatic evolution and weak system-environment coupling.
In addition, they are not in the standard Lindblad form and they do not {\it a priori} conserve the positivity  of the density matrix. For this reason, particular attention must be paid to the parameters used in the numerical simulations since they can lead to inaccurate results in the non-adiabatic case and strong coupling regime.

\subsection{Secular Approximation}
\label{sec:secular}

In Eqs. (\ref{eq:Eq1}) and (\ref{eq:Eq2}), we have included all the contributions up to the second order in $\alpha$ and $V$  contrary to the usually adopted secular approximation. \cite{cohen-tannoudji}
In the interaction picture, the non-secular terms oscillate rapidly with respect to the relaxation-dephasing dynamics generated by the environment and thus their contribution averages to zero.
Even though this approximation does not lead to problems  in many cases, we show that it is inadequate in describing adiabatic evolution.

To compare the results with and without these terms, we explicitly write the master equations corresponding to Eqs.  (\ref{eq:Eq1}) and (\ref{eq:Eq2}) but obtained with the secular approximation as
\begin{widetext}
\begin{eqnarray}
 \dot{\rho}_{gg}^{\rm sec} &=& -2 \Im \mbox{m}(w_{ge}^* \rho_{ge})+S(\omega_0 ) \left|m_2\right|{}^2-[S(-\omega_0 )+S(\omega_0 )]\left|m_2\right|{}^2 \rho_{gg} \nonumber\\ 
   &+&2\frac{2 S(0)-S(-\omega_0 )-S(\omega_0 )}{\omega_0 }\left\{\Im \mbox{m}\left(m_2\right) \Im \mbox{m}\left(w_{ge}\right)+\Re \mbox{e}\left(m_2\right)
   \Re \mbox{e}\left(w_{ge}\right)\right\} m_1 \rho _{gg}  \nonumber\\ &&
   -2 m_1\frac{S(0)-S(\omega_0)  }{\omega_0 }\left\{\Im \mbox{m}\left(m_2\right)
   \Im \mbox{m}\left(w_{ge}\right)+\Re \mbox{e}\left(m_2\right) \Re \mbox{e}\left(w_{ge}\right)\right\} ,
  \label{eq:Eq1_sec}
\end{eqnarray}
\end{widetext}
and
\begin{widetext}
\begin{eqnarray}
 \dot{\rho}_{ge}^{\rm sec} &=& i w_{ge} (2 \rho_{gg}-1) +i (w_{ee}-w_{gg}) \rho_{ge} + i \omega_0 \rho_{ge} - [\frac{1}{2}S(-\omega_0 ) \left|m_2\right|{}^2
 +\frac{1}{2} S(\omega_0 ) \left|m_2\right|{}^2+2 S(0) m_1^2] \rho_{ge} \nonumber\\ 
 &+&\frac{2 S(0)-S(-\omega_0 )-S(\omega_0 )}{\omega_0 } ( 2 m_2 w_{eg}+w_{ge} m_2^*  ) m_1  \rho_{ge} .
\label{eq:Eq2_sec}
\end{eqnarray}
\end{widetext}
Notice that, in the time-independent basis (i.e. $w_{kl} \equiv 0$), these reduce to the usually adopted Bloch equations for a non-driven system
\begin{eqnarray}
 \dot{\rho}_{gg}^{\rm B} &=& \Gamma_{\downarrow} \rho_{ee} - \Gamma_{\uparrow} \rho_{gg}, \nonumber\\ 
 \dot{\rho}_{ge}^{\rm B} &=&  i \omega_0 \rho_{ge} - \Gamma_{ge} \rho_{ge},
\end{eqnarray}
where $\Gamma_{\downarrow}=S(\omega_0 ) \left|m_2\right|^2$, $\Gamma_{\uparrow}=S(-\omega_0 ) \left|m_2\right|^2$, and $\Gamma_{ge}=\left[\frac{1}{2}S(-\omega_0 ) \left|m_2\right|{}^2+\frac{1}{2} S(\omega_0 ) \left|m_2\right|{}^2+2 S(0) m_1^2\right]$.
These equations have immediate interpretation in terms of relaxation and dephasing processes.

\subsection{Adiabatic quasi-stationary solutions for the ground state and the role of secular approximation}
\label{sec:quasi-stationary}

In absence of system-environment interaction, i.e., $S(\omega)=0$, Eqs.~(\ref{eq:Eq1}) and~(\ref{eq:Eq2}) become 
the standard  von Neumann equations for the density matrix
\begin{eqnarray}
 \dot{\rho}_{gg} &=& -2 \Im \mbox{m} ( w_{ge}^* \rho_{ge}), \nonumber\\
 \dot{\rho}_{ge} &=& i w_{ge}(2 \rho_{gg}-1)+i (w_{ee}-w_{gg})\rho_{ge}+i \omega_0 \rho_{ge}.
 \label{eq:idealEqs}
\end{eqnarray}
Since we are interested in the evolution in the quasi-stationary limit for adiabatic evolution ($\alpha \ll 1$), we look for the solution of $\dot{\rho}_{gg} =0$ and $\dot{\rho}_{ge} =0$.
We have 
\begin{eqnarray}
 \rho_{gg} &=& 1+ O(\alpha^2), \nonumber\\
 \rho_{ge} &=&-\frac{w_{ge}}{\omega_0} +O(\alpha^2).
 \label{eq:idealsolution}
\end{eqnarray}

In general, Eqs.  (\ref{eq:Eq1}) and (\ref{eq:Eq2}) can be integrated only numerically. 
However, we can obtain analytical information if we exploit the adiabatic limit. 
As above, the correction to the ground-state population $\rho_{gg}$ is of order $\alpha^2$ 
and the off-diagonal element $\rho_{ge}$ scales as $\alpha$.

In the following, we consider first the finite temperature case, in which we assume that the excitation rates are exponentially small with respect to the relaxation ones; for example, $\Gamma_{ge}=\Gamma_{eg} \exp{[-\hbar \omega_0/(k_B T)]}$, where $T$ is the temperature of the bath.
To the lowest order in $ \alpha$ and in the quasi-stationary limit, the solution for $\rho_{gg}$ has a simple expression
\begin{equation}
\rho_{gg} = \frac{\Gamma_{eg}}{\Gamma_{eg}+\Gamma_{ge}}  +O(\alpha^2) \approx 1 - e^{-\frac{\hbar \omega_0}{k_B T}}  +O(\alpha^2).
\label{eq:solutionfiniteTgg}
\end{equation}
The system remains in the ground state in the linear order in $\alpha$, apart from leakage due to the finite temperature environment.

Within the same approximation and using the above result, the equation for the off-diagonal term reads
\begin{eqnarray}
  \dot{\rho}_{ge} &=&(i \omega_0  -2 S(0) m_1^2) \Delta_{ge,1} \nonumber\\ 
  &+ &i \Im \mbox{m}\left(m_2\right) m_2 S(\omega_0) \Re \mbox{e}(\Delta_{ge,2}) \nonumber\\ 
  & -&i \Re \mbox{e}\left(m_2\right) m_2 S(\omega_0) \Im \mbox{m}(\Delta_{ge,2})=0, \nonumber\\ 
 \label{eq:Delta_eq}
\end{eqnarray}
where we have defined 
\begin{eqnarray}
 \Delta_{ge,1} &=& \frac{\left(1-2 e^{-\frac{\hbar \omega_0}{k_B T}}\right) w_{ge}}{\omega_0 }+\rho _{ge}, \nonumber\\
 \Delta_{ge,2} &=& \left(1+e^{-\frac{\hbar \omega_0}{k_B T}}\right) \rho_{ge}+ \frac{\left(1-e^{-\frac{\hbar \omega_0}{k_B T}}\right)  w_{ge}}{\omega_0 }. 
\end{eqnarray}
At low but finite temperatures, the solution of Eq. (\ref{eq:Delta_eq}) is 
\begin{equation}
\rho_{ge} = -\frac{w_{ge}}{\omega_0} (1-2 e^{-\frac{\hbar \omega_0}{k_B T}}) ~ .
\label{eq:solutionfiniteTge}
\end{equation}
Notice that, in the zero-temperature limit, solutions  in Eqs. (\ref{eq:solutionfiniteTgg}) and (\ref{eq:solutionfiniteTge}) coincide with the the environment-free solutions (\ref{eq:idealsolution}): the ground-state evolution is robust and not influenced by the zero-temperature environment.

This can be explained in a simple way. \cite{pekola09}
The effective Hamiltonian in the absence of system-environment coupling is $\tilde{H}_S+ \hbar w$. 
In the adiabatic limit, the second term can be treated as a perturbation and the correction to the 
eigenstates of $H_S$ can be calculated. This basis is usually called superadiabatic. \cite{whitney05}
Up to the linear order in $\alpha$, the superadiabatic ground state is
$|g^\prime \rangle = (|g\rangle - w_{ge}^*/\omega_0 |e\rangle)$.
In the adiabatic evolution, if the system starts in the ground state $|g^\prime(0)\rangle$, we can assume that it remains in the eigenstate $|g^\prime(t)\rangle$ and the corresponding density matrix is $\rho^\prime= |g^\prime(t)\rangle \langle g^\prime(t)|$.
Thus, the density matrix elements in the $H_S$ basis,
$\rho_{gg}=\langle g | \rho^\prime | g\rangle$ and $\rho_{ge}=\langle g | \rho^\prime | e\rangle$ satisfy Eq. (\ref{eq:idealsolution}).

\begin{figure*}
    \begin{center}
     \hspace{1cm}
    \includegraphics[height=5cm,width=5cm ]{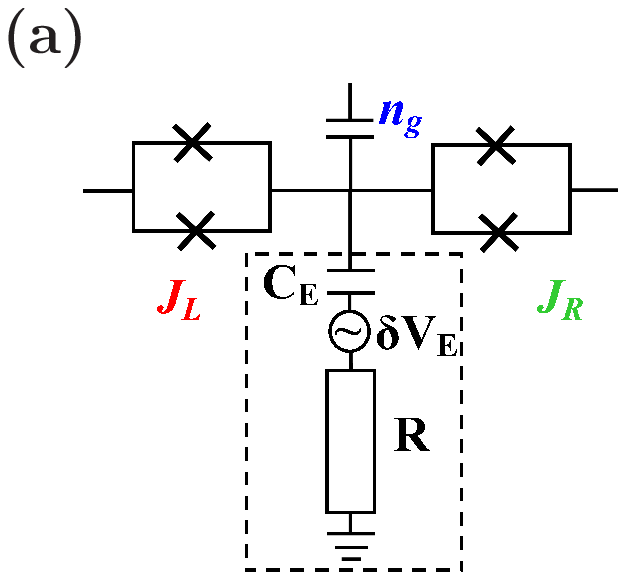} 
    \includegraphics[scale=0.7]{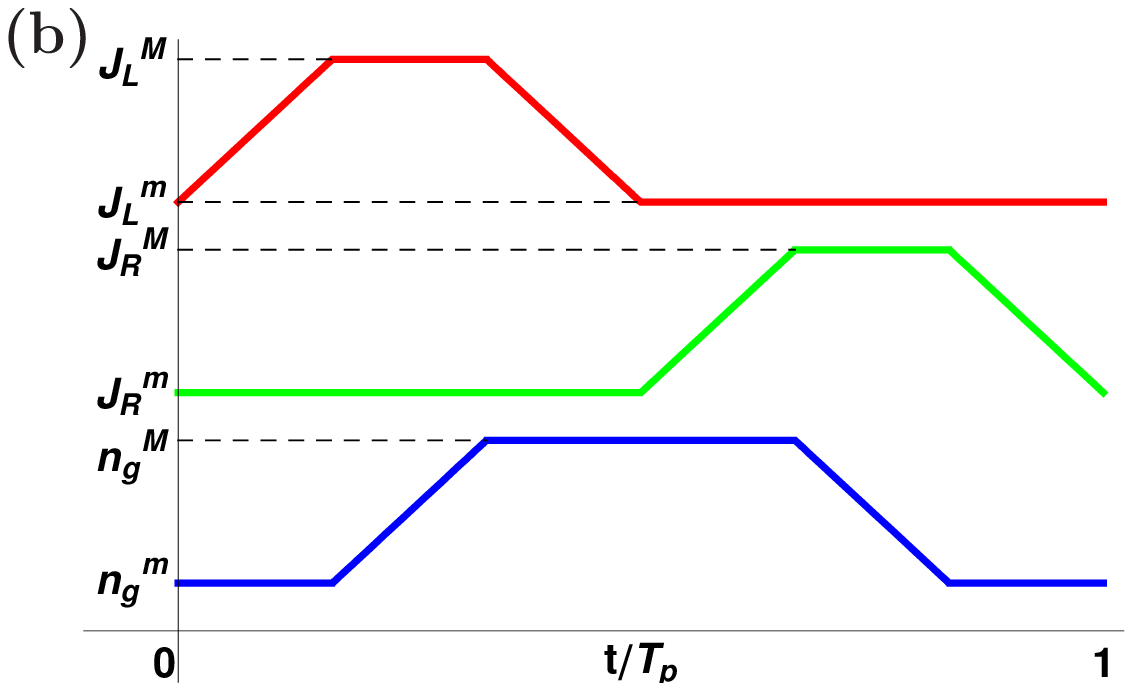}
    \end{center}
    \caption{(Color online) (a) Circuit scheme of  the Cooper pair sluice (physical realization of an adiabatic two level system). 
    The time-dependent control parameters are $J_R$, $J_L$, and $n_g$. The phase difference across the device is denoted by $\varphi$.
    The circuit in the dashed box is used to engineer the environmental noise characterized and modeled by a resistor $R$, a voltage $\delta V_E$ and coupled to the system with an effective capacitance $C_E$.
    (b) The time-dependence of the parameters $\{J_L,J_R,n_g\}$ in a pumping cycle.}
    \label{fig:scheme}
\end{figure*}

This result helps to understand the robustness of ground-state evolution subject to zero-temperature noise.
In this regime, the environment cannot produce transitions towards the excited state $|e^\prime(t)\rangle$ but induces only  relaxation to the ground state $|g^\prime(t)\rangle$. 
Since the system starts from the ground state and evolves adiabatically, it remains always in the ground state of $\tilde{H}_S+ \hbar w$ and the environment has no influence on its dynamics.
In this framework, it is clear that neglecting the $w(t)$ terms leads to relaxation with respect to the basis of $\tilde{H}_S$ and hence to different dynamics.

The above result demonstrates the importance of the choice of the proper master equation.
To emphasize this point further, we use a similar approach to solve the master equation arising from the secular approximation [Eqs. (\ref{eq:Eq1_sec}) and (\ref{eq:Eq2_sec})].
In particular, in the zero-temperature limit and up to the order of the product of $\alpha$ and the square of the system-environment coupling, we obtain
\begin{eqnarray}
\rho_{gg}^{\rm sec} &=& 1, \nonumber\\ 
\rho_{ge}^{\rm sec} &=& -\frac{2 i w_{ge}}{ 2 i \omega_0 -  \Gamma} ,
 \label{eq:secular-stationary-limit-zeroT}
\end{eqnarray}
where $\Gamma= S(\omega_0 ) \left|m_2\right|{}^2+4 S(0) m_1^2$ and the corrections are of order of the product of the adiabatic parameters and the square of the system-environment coupling .
The expression of the diagonal term is not influenced by the secular approximation, but the off-diagonal one depends on the coupling between the system and the environment.
This is in contrast with both the physical picture discussed above and analytical results in Eq. (\ref{eq:solutionfiniteTge}). This difference leads to different expectation values for physical observables.

We can obtain analytical information also in the opposite regime in which the evolution is fast enough that non-adiabatic transitions determine the dynamics of the system.
In this limit, neglecting the relaxation terms in Eqs.~(\ref{eq:Eq1}) and~(\ref{eq:Eq2}), we obtain again Eq.~(\ref{eq:idealEqs}). 
The system is initially in the ground state with $\rho_{gg}(0)=1$ and $\rho_{ge}(0)=0$.
We expect that, due to sequential non-adiabatic excitations, in the quasi-stationary limit both the ground and the excited state are equally populated, i.e., $\rho_{gg}=\rho_{ee}=1/2$. 
With this assumption, the differential equation for $\rho_{ge}$ in (\ref{eq:idealEqs}) can be solved to obtain the quasi-stationary solution
\begin{eqnarray}
\rho_{gg} &=& 1/2, \nonumber\\ 
\rho_{ge} &=& 0
 \label{eq:non-adiabaticSolution}
\end{eqnarray}
apart from corrections of the second order in the system-environment coupling.
The system undergoes sequential transitions from the ground to the excited state and vice versa forming a fully mixed state.

Thus, in the two limits in which relaxation dominates non-adiabatic transitions and non-adiabatic slight transitions dominate relaxation, it is possible to find an analytical quasi-stationary solution which correctly describes the system behavior [Eqs.~(\ref{eq:idealsolution}) and~(\ref{eq:non-adiabaticSolution}), respectively].
No such solution is available in the intermediate regime in which relaxation and non-adiabatic transitions occur on the same timescale.

\section{Application to Cooper pair pumping}
\label{sec:physical_model}

In the previous section, we showed that ground-state evolution can be robust against environmental noise.
The information about the dynamics of the system is contained in the off-diagonal term of the density matrix, see Eq. (\ref{eq:solutionfiniteTge}).
Thus to test our theoretical model, it is natural to focus on a physical observable which depends on $\rho_{ge}$.
An interesting candidate for the observable is the pumped charge in a superconducting circuit.
Recently, it has been subject to intense study both theoretically \cite{aunola03} and experimentally \cite{mottonen08}
because of its connection to geometric phases, \cite{aunola03,mottonen06, leone08} and its potential application in quantum  metrology. \cite{niskanen03, niskanen05, vartiainen07}

In particular, we consider the Cooper pair sluice \cite{niskanen03} shown in Fig. \ref{fig:scheme}. 
It consists of a single superconducting island, coupled to superconducting leads via two  superconducting quantum interference devices (SQUIDs).
The SQUIDs operate as Josephson junctions whose critical currents can be tuned by magnetic fluxes.
The electrostatic potential on the island can be controlled by varying a gate voltage, $V_g$, and there is a constant superconducting phase difference, $\vf=\vf_R+\vf_L$ between the two leads.
The operator $n_k=-i \partial_{\varphi_k} $ ($k=L, R$) represents the Cooper pair number operator of the $k$-th SQUID.
In the absence of noise, the Hamiltonian of the sluice can be expressed as \cite{pekola09}
\begin{equation}
H_S =E_C(n-n_g)^2 -J_L \cos (\frac{\varphi}{2} - \theta) -J_R \cos (\frac{\varphi}{2} + \theta)
 \label{eq:H_S}.
\end{equation}

Here $\theta=(\vf_R-\vf_L)/2$ and $n= -i \partial_\theta$ are the operators for the superconducting phase and the number operator of excess Cooper pairs on the island.
The Josephson couplings to the left and right lead are denoted by $J_L$ and $J_R$, respectively, $n_g=C_gV_g/(2e)$ is the normalized gate charge,  and  $E_{C}= 2e^2/C_{\Sigma}$ is the charging energy of the sluice; $C_g$ is the gate capacitance and $C_{\Sigma}$ the total capacitance of the island. For further convenience, we define the deviation of the gate charge from the degeneracy point as $\delta n_g= n_g-1/2$.

The average-value given by the current through the $k$-th SQUID is  $I_k= -2 e \mbox{Tr}(\dot{\rho}_{\rm tot} n_k)$ where the trace is over the degrees of freedom of both the system and the environment, and $\rho_{\rm tot}$ is the total density matrix of the system and the environment.
Using the von Neumann equation for $\rho_{\rm tot}$, we obtain
\begin{eqnarray}
  I_k &=& \frac{2 i e }{\hbar}i\mbox{Tr}(\rho_{\rm tot} [n_k, H_S(t)]) +\mbox{Tr}(\rho_{\rm tot} [n_k, H_E]) \nonumber\\ 
  && +\mbox{Tr}(\rho_{\rm tot} [n_k, V]).
  \label{eq:current_definition}
\end{eqnarray}
Writing the complete trace in terms of partial traces and tracing out only the environmental degrees of freedom,  we obtain  $\mbox{Tr}_S(\rho ~[n_k, H_S(t)])$ which is the usual definition of the current for a closed system. \cite{mottonen06}
The second term gives no contribution since $n_k$ and $H_E$ commute.
The third term is the current directly induced by the environment.
Notice that, the last contribution is present only if $[n_k, V] \neq 0$. In all the other cases, we can use the 
standard definition of current operator: $I_k = \frac{2 e i }{\hbar} [n_k, H_S(t)]$.

\begin{figure*}
    \begin{center}
    \includegraphics[width=7.5cm]{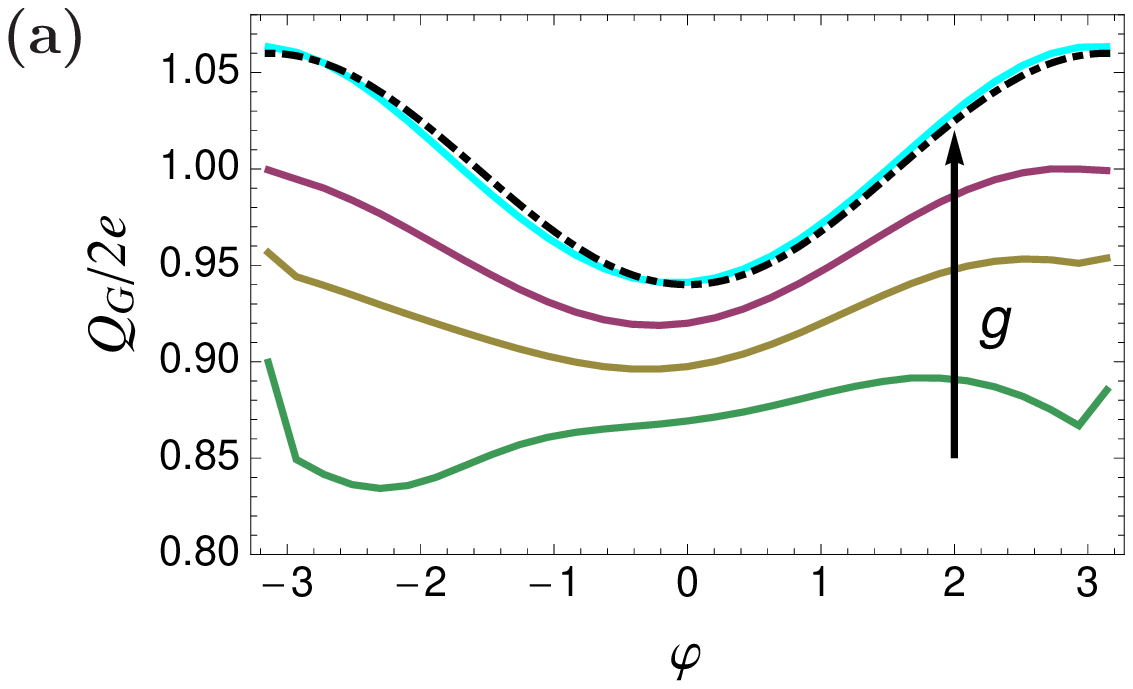}
    \includegraphics[width=7.5cm]{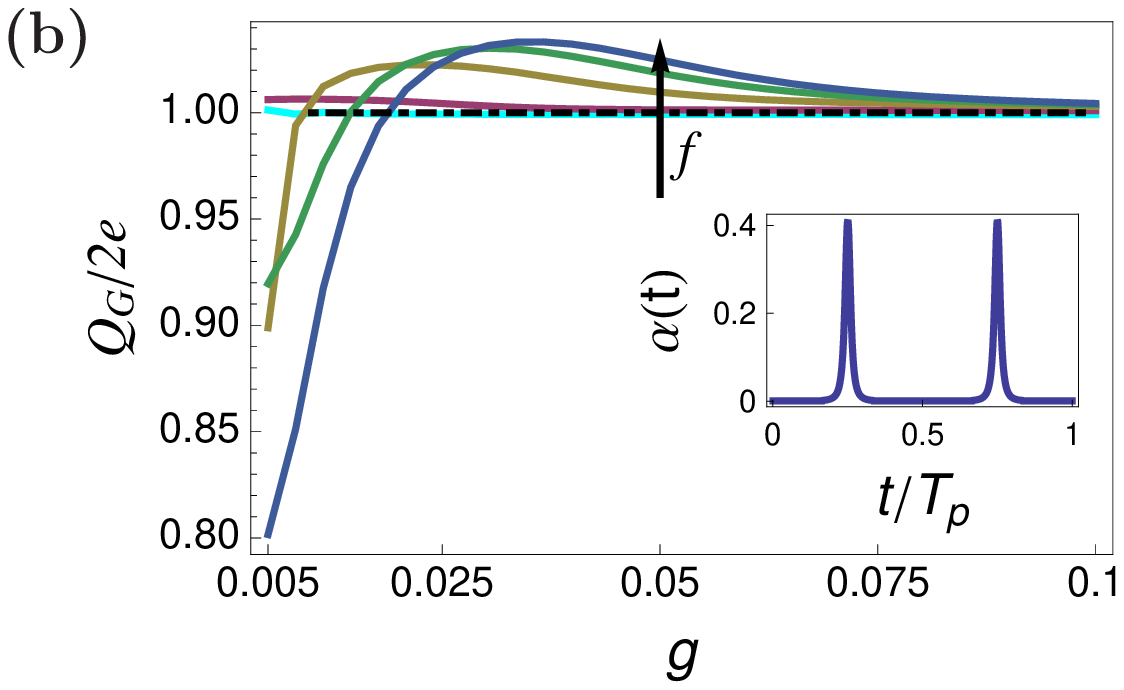}
    \end{center}
    \caption{(Color online) Pumped charge per cycle in a Cooper pair sluice for zero-temperature environment (a) as a function of phase $\varphi$ with coupling constant $g=C_E/C_\Sigma=0.01, 0.0125, 0.015, 0.1$ from bottom to top for $f=1/T_p=75$ MHz, and (b) as function of coupling $g$ with $\varphi=\pi/2$ and at frequencies $f=10, 25, 50, 65, 75$ MHz from bottom  to top in the region in which the arrow is plotted.
    In the numerical simulations, the parameters are $J_i^M/E_{\rm C}=0.1$, $J_i^m/J_i^M=0.03$ (with $i=L,R$), $\delta n_g^M=
    -\delta n_g^m=0.3$, $E_{\rm C}/k_{\rm B}=1$ K [$E_{\rm C}/(2\pi\hbar)= 21$ GHz], $R=300$ k$\Omega$, environment temperature $T=0$, 
    $S(\omega_0)=2 \hbar \omega_{0}R$, $S(-\omega_0)=0$, and $S(0) = 2k_{\rm B}T_0R$ with $T_0=0.1$~K.
    In inset we show the time evolution of the adiabatic parameter $\alpha(t)$ for $f=75$ MHz. %%  here
    }
    \label{fig:Q}
\end{figure*}

We study the case in which the noise is due to gate voltage fluctuations coupled capacitively to the sluice island. \cite{makhlin01}
The charging part of the Hamiltonian in Eq. (\ref{eq:H_S}) has the form $ E_C [n-C_g/(2e) V_g ]^2$.
If an additional noisy gate potential $\delta V_E$ is coupled to the system through a capacitance $C_E$, we can use the above expression to write the Hamiltonian describing the interaction between the system and the environment.
At the lowest order in $\delta V_E$ the Hamiltonian can be written as $V = - 2 e g n \otimes \delta V_E(t)$, where $g=C_E/C_\Sigma$ is the coupling constant between the system and environment.
Notice that since $[n_k,V]=0$, the current operator of the $k$th SQUID is determined only by the first term in Eq. (\ref{eq:current_definition}). 
Its average value reads
\begin{equation} 
	I_k=(\rho_{gg}I_{k,gg}+\rho_{ee}I_{k,ee})+ 2 \Re \mbox{e}{(\rho_{ge}I_{k,eg})},
\end{equation}
where $I_{k,nm} = \langle n | I_k |m\rangle$ with $m,n=g,e$.
The last term is the geometric contribution corresponding to the pumped charge
\begin{equation} 
	Q_k^G=\int_0^{T_p} 2 \Re \mbox{e}{(\rho_{ge}I_{k,eg})} dt.
	\label{eq:Q_G}
\end{equation}

If $E_C \gg \mbox{max}\{J_L,J_R\}$ and $n_g \approx 1/2$, only the two lowest charge states are important for the dynamics and we can adopt the two-state approximation.
Let $|0\rangle$ and $|1\rangle$ denote the states with no and one excess Cooper pair on the island, respectively.
Thus the coupling between the sluice and charge noise has the form 
$V =  e g \s_z \otimes \delta V_E(t)  $, where $\s_z=|0\rangle\langle 0|-|1\rangle \langle1|$ a part terms proportional to identity.

\subsection{Gate charge noise}
\label{sec:gate_charge_noise}

With the above notations and in this approximation, 
we can express all the relevant quantities in terms of the control parameters $J_L$, $J_R$ and $n_g$ as
\begin{eqnarray}
  E_{12} &=& \frac{1}{2} \sqrt{J_L^2+J_R^2 + 2 J_L J_R \cos \varphi}, \nonumber \\
  \gamma &=& \arctan \left(\frac{J_R-J_L}{J_R+J_L} \tan \frac{\varphi}{2} \right), \nonumber \\
  \eta   &=& \frac{ \delta n_g}{\sqrt{\delta n_g^2 + (\frac{E_{12}}{E_C}})^2},  \nonumber \\
  \omega_0 &=& \frac{2 E_{12}}{\hbar \sqrt{1-\eta^2}}, %\nonumber \\
\label{eq:parameters}
\end{eqnarray}
and the instantaneous eigenstates assume the form
\begin{eqnarray}
 |g\rangle &=& \frac{1}{\sqrt{2}}( \sqrt{1-\eta} |0\rangle + e^{-i \gamma} \sqrt{1+\eta} |1\rangle  ), \nonumber\\
 |e\rangle &=& \frac{1}{\sqrt{2}}(\sqrt{1+\eta} |0\rangle - e^{-i \gamma}\sqrt{1-\eta} |1\rangle  ) .
\end{eqnarray}
Further, the matrix elements of $Y$ ($V= Y\otimes X$) read $\hbar m_1 = -g \eta$ and $\hbar m_2 = g \sqrt{1-\eta^2}$.

\begin{figure*}
	\centering
		\includegraphics[width=7.5cm]{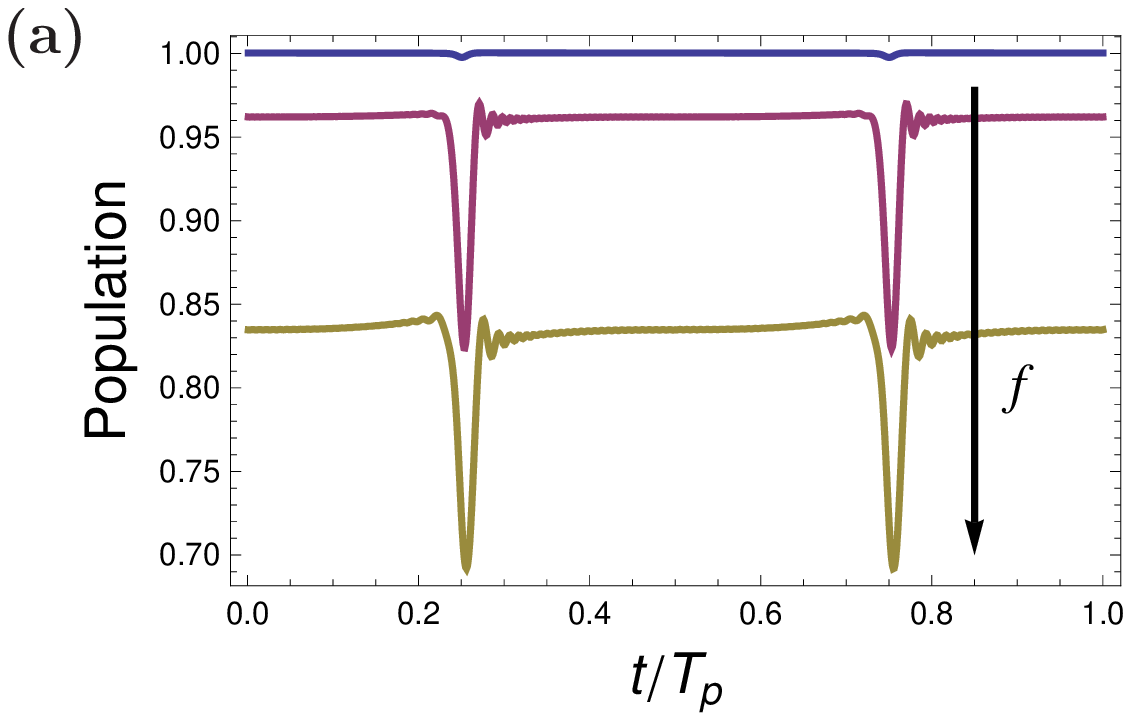}
		\includegraphics[width=7.5cm]{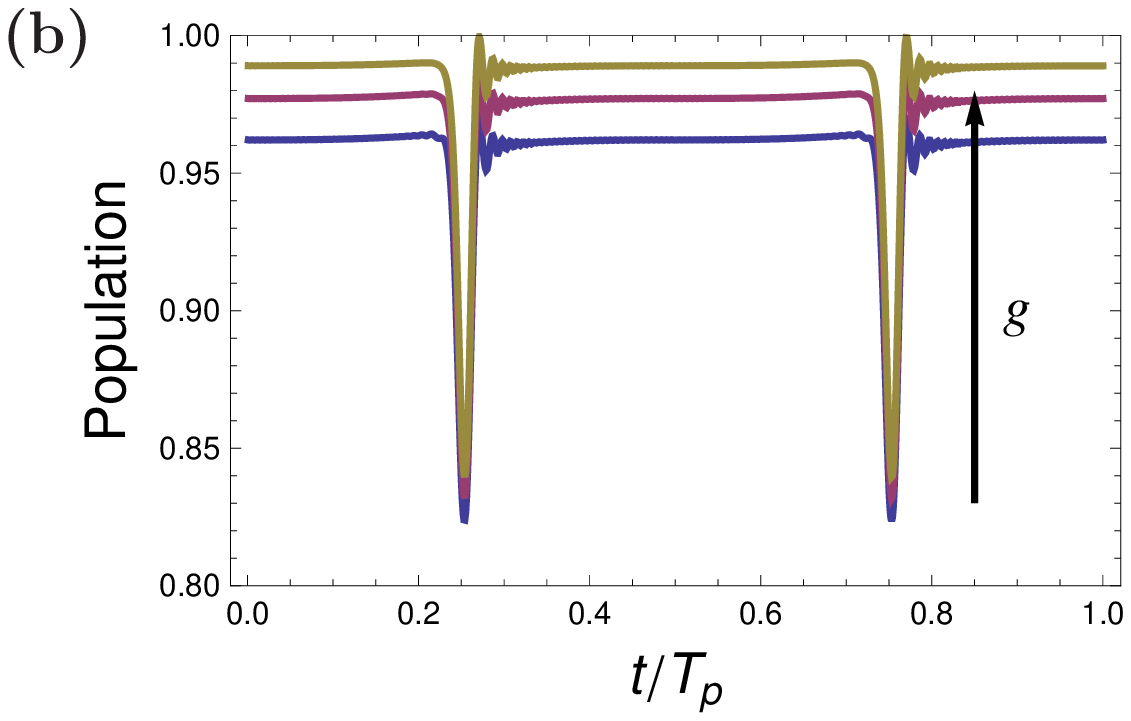}
	\caption{(Color online) Population of the ground state of $\tilde{H}_S+\hbar w$ in the quasi-stationary evolution.
	(a) Evolution with $g=0.01$ and  $f=10, 75, 100$ MHz from top to bottom.
	(b) Evolution in presence of environmental noise for a non-adiabatic $75$ MHz cycle with $g=0.01, 0.0125, 0.015$ from bottom to top.	The other parameters are the same as in Figure \ref{fig:Q}.
	}
	\label{fig:superad_basis}
\end{figure*}

The $w$ matrix elements we need are 
\begin{eqnarray}
 w_{gg} &=& -\frac{1}{2}( 1+\eta) \dot{\gamma} \nonumber\\
 w_{ee} &=& -\frac{1}{2}( 1-\eta) \dot{\gamma} \nonumber\\
 w_{ge} &=& \frac{1}{2} ( \sqrt{1-\eta^2} \dot{\gamma}- i \frac{\dot{\eta}}{\sqrt{1-\eta^2}}  )
 \label{eq:W}
\end{eqnarray}

Using $I_k = 2e \partial_{\varphi_k}  H_S   /\hbar$, \cite{mottonen06} the current operators are $I_L = (2e /\hbar) J_L \sin(\frac{\varphi}{2} - \theta)$ and $I_R = (2e /\hbar) J_R \sin(\frac{\varphi}{2} + \theta)$.
We can write the current operators restricted to the $\{ |0\rangle, |1\rangle \}$ basis using the formulas $e^{i \theta} = |1\rangle \langle 0|$ and $e^{-i \theta} = |0\rangle \langle 1|$.
Inserting them in Eq. (\ref{eq:Q_G}), the charge pumped through each junction reads
\begin{eqnarray}
 Q_L^G &=& \frac{2 e }{ \hbar} \int_0^{T_p}dt  J_L \Big( \eta  \Re \mbox{e}\left(\rho _{ge}\right) \sin \left(\gamma +\frac{\varphi}{2}\right) \nonumber\\
   && -\cos \left(\gamma +\frac{\varphi }{2}\right) \Im \mbox{m}\left(\rho _{ge}\right) \Big) \nonumber\\
 Q_R^G &=& \frac{2 e }{ \hbar} \int_0^{T_p} dt J_R\Big(\cos \left(\gamma -\frac{\varphi }{2}\right) \Im \mbox{m}\left(\rho
   _{ge}\right) \nonumber\\
   && -\eta  \Re \mbox{e}\left(\rho _{ge}\right) \sin \left(\gamma -\frac{\varphi }{2}\right)\Big) .
   \label{eq:charge_eq}
\end{eqnarray}

We define the average pumped charge as $Q^G \equiv (Q_L^G+Q_R^G)/2$ and charge conservation implies that $Q^G= Q_L^G=Q_R^G$.

In our simulations the spectral density is taken to be ohmic such that $S(\omega) =2 \hbar \omega R /[1-e^{-\hbar \omega/(k_B T)}]$ where $R$ is the effective resistance of the environmental noise.
For finite temperatures, we have $S(-\omega)=e^{-\hbar \omega/(k_B T)} S(\omega)$ and $S(0)=2 k_B T_0 R$, where $T_0$ is the effective temperature of the dephasing process.
We note that the master equations presented in this paper are also valid for other types of environment spectra.

\subsection{Numerical Results}
\label{sec:numerical}

The pumping cycle taken into consideration is shown in Fig.~\ref{fig:scheme}(b) and the corresponding evolution of the quantum system is obtained by numerical integration of Eqs. (\ref{eq:Eq1}) and (\ref{eq:Eq2}).
The physical observable here is the pumped charge through the junction in the stationary regime which is reached after several sequential pumping cycles, i.e., after the initial transient is over.
In fact, during the first cycles the pumped charge oscillates  due to simultaneous effects of non-adiabatic excitation and environmental relaxation and it takes several cycles to stabilize to the stationary values which are measured in the experiments.

In Fig.~\ref{fig:Q}(a), the pumped charge is shown as a function of the phase $\varphi$ across the device $\varphi$ and for different values of  the coupling strength $g$ while the pumping frequency is kept constant. 
The ideal pumped charge in the absence of noise \cite{niskanen03} is also shown and it coincides with the numerical  result in the strong coupling limit.
For weak coupling to the environment, the pumped charge differs substantially from the ideal one but an increment of the system-environment coupling resumes the ideal pumping.

In Fig.~\ref{fig:Q}(b), we present the numerically simulated pumped charge as a function of $g$, for different pumping frequencies.
For an adiabatic loop (e.g. $f=1/T_p=10$ MHz), the pumped charge is close to the ideal value. This confirms the first analytical results: in the adiabatic limit and $T=0$, the environmental noise does not influence the evolution.
For faster pumping frequencies, the behavior is more complex because it is determined by the competing effects of non-adiabatic transitions and environmental relaxation.
When the non-adiabatic transitions dominate over the relaxation process the predicted pumped charge is smaller than the ideal one. This is due to the fact that the excited state $|e^\prime\rangle$ of $\tilde{H}_S+\hbar w$ is populated during the evolution and it pumps in the opposite direction with respect to the ground state $|g^\prime\rangle$.
Increasing the coupling we enter the regime in which the relaxation process dominates. Here, the system is forced to remain in the ground state $|g^\prime\rangle$  and we restore the ideal pumping. 
Thus, the numerical simulations confirm the analytical results obtained in Sec. \ref{sec:quasi-stationary} for different value of $\alpha/g$ ratio.

\begin{figure*}
	\centering
		\includegraphics[width=7.5cm]{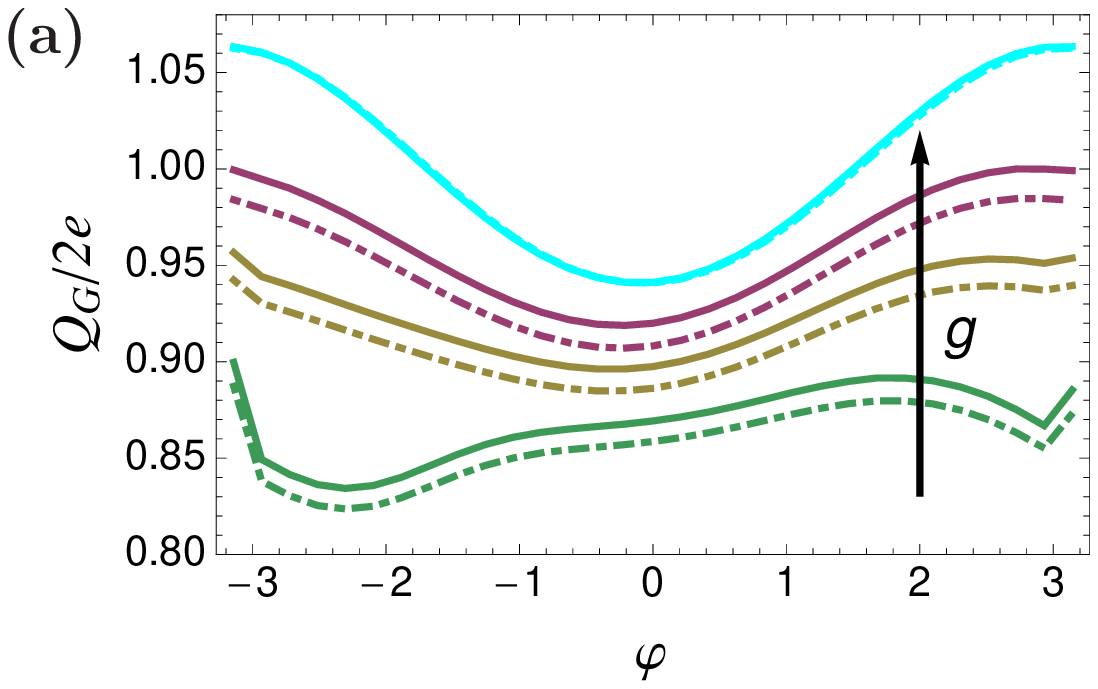}
		\includegraphics[width=7.5cm]{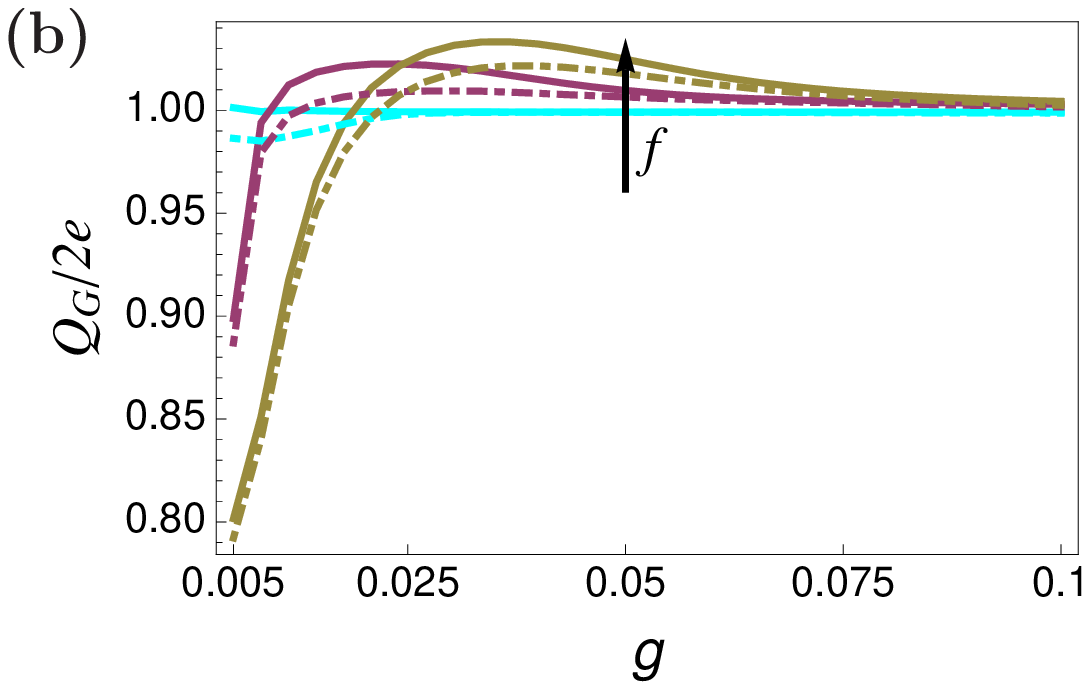}
	\caption{(Color online) Pumped charge per cycle for finite-temperature environments (dashed curves) in comparison to the zero-temperature environment case (solid lines):  (a) as a function of phase $\varphi$ for $f=75$ MHz and  $g=0.01, 0.0125, 0.015, 0.1$ from bottom to top and (b) as a function of the coupling strength $g$ with $\varphi=\pi/2$ and frequency $f=10,50,75$ MHz from bottom to top. The other parameters are as in Fig. \ref{fig:Q} except for $T=0.03$ K for the finite temperature simulations.}
	\label{fig:Q_finite_temp}
\end{figure*}

To check this interpretation of the numerical results, we calculate the population  of $|g^\prime\rangle$ during the pumping cycles for different environmental coupling strengths.
This is obtained by solving Eqs.~(\ref{eq:Eq1}) and~(\ref{eq:Eq2}) numerically  and projecting the obtained density matrix into the  $\{ |g^\prime\rangle, |e^\prime\rangle \}$ basis.
Figure~\ref{fig:superad_basis}(a) shows how increasing the frequency  leads to a leakage from the ground $|g^\prime\rangle$ to the excited $|e^\prime\rangle$ state.
The figure shows that increasing the coupling strength from the region $g \ll \alpha \ll 1$ to $\alpha \ll g \ll 1$, the system begin to remain in the ground state $|g^\prime\rangle$, thus restoring the evolution expected in the adiabatic limit without noise.
This is a generic feature and we have chosen $75$ MHz pumping frequency to illustrate the phenomenon.

Even if the convergence to the ideal evolution has strong physical motivation, we stress that the details of the numerical simulations in the extreme regime must be taken with care.
In fact, in the intermediate regime both the analytical and numerical approaches fail to predict the correct behavior of the pumped charge.
In this regime, the numerical simulations in Fig.~\ref{fig:Q}(b) are inaccurate and present an excess of pumped charge with respect to that in the adiabatic limit.
This unphysical prediction is due to the fact that the master equation does not guarantee  a priori the positivity of the density matrix and that we have truncated at the first order in $\alpha$ and $g$ in the derivation of the master equation. 
Strictly speaking the $g$ and $\alpha$  used are always small compared to $1$ [see  Fig.~\ref{fig:Q}(b)] and thus, at least formally, they can be used as expansion parameters.
Unfortunately, such analysis gives only an estimate of the corrections we are neglecting and proper analysis should include the higher order terms. \cite{salmilehto10}

However, there are indications that the features in the $\alpha$ and $g$ regimes presented  are realistic.
The convergence to the ideal solution predicted by the analytical analysis is present in the simulations for all the range of parameters used.
It seems unlikely that such common behavior, which is realistic for small $\alpha$ and $g$, is a numerical artifact.
The violation of positivity, which has a role in the intermediate regime, becomes less important for the maximum value of $g$ presented. In fact, the analysis performed confirms that in these extreme regimes the positivity tends to be restored.

\subsection{Finite Temperature Pumping}

In Fig.~\ref{fig:Q_finite_temp}, we compare Cooper pair pumping with zero- and finite-temperature environments.
We observe that the features of the zero-temperature simulation are still present although, at finite-temperature,  we obtain in general a smaller pumped charge with the same parameter values. 
This is expected since the presence of finite-temperature environment leads to an additional leakage 
from the ground state to the excited state, leading to a reduced pumped charge since the adiabatically pumped charge in the excited state is equal in magnitude but opposite in sign to that in the ground state. \cite{mottonen06}

\subsection{Pumped charge with secular approximation and charge conservation}
\label{sec:charge-conservation}

All the results presented above depend on whether one includes in the master equation the non-secular terms  and the combined effect of the environment and the driving  or not .
From the analytical calculation presented in Sec. \ref{sec:secular}, we know that, in the quasi-stationary limit, the off-diagonal elements of density matrix  within the secular approximation [Eq. (\ref{eq:secular-stationary-limit-zeroT})] have a different form compared with the full expression [Eq. (\ref{eq:solutionfiniteTge})]. Thus we expect this to lead to different values of physical observables.
To compare the two methods, the pumped charge is again a privileged candidate since it depends directly on the off-diagonal matrix element $\rho_{ge}$.

\begin{figure}
	\centering
		\includegraphics[width=7.5cm]{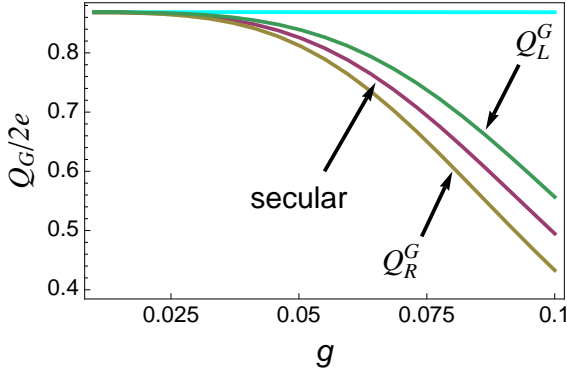}
	\caption{(Color online) Pumped charge obtained using different master equations presented for $f=10$ MHz cycle. 
	The straight line is the solution of the full master equation (\ref{eq:Eq1}) and
	 (\ref{eq:Eq2}), the solid lines are the solutions of the master equation with secular approximation (\ref{eq:Eq1_sec}) and (\ref{eq:Eq2_sec}).
	 In the latter case, the central solid line is the average pumped charge while the other two lines are the pumped charges through the left and right junctions.
	 Here, the parameter cycle is strongly asymmetric with 
   $J_L^M/E_{\rm C}=0.1$, $J_L^m/J_L^M=0.006$, 
   $J_R^M/E_{\rm C}=0.2$, $J_R^m/J_R^M=0.04$, 
   $\delta n_g^M=0.4$ $\delta n_g^m=-0.03$, $T=0$, and the other parameters are as in Fig.
    \ref{fig:Q}.	 
	 }
	\label{fig:Charge_Conservation}
\end{figure}

In the quasi-stationary limit, within the secular approximation, at zero temperature, and for weak system-environment coupling, we obtain from Eq. (\ref{eq:secular-stationary-limit-zeroT})
\begin{equation}
\rho_{ge}^{\rm sec} \approx  -\frac{ w_{ge}}{ \omega_0} \left(1-\frac{i \Gamma }{2 \omega_0} -\frac{\Gamma ^2}{4 \omega_0 ^2} \right) = \rho_{ge}- \delta \rho,
\label{eq:rho_sec}
\end{equation}
where $\delta \rho$ denotes the correction with respect to the full solution. 
From this expression it results that $\rho_{ge}^{\rm sec}$ is damped because of the interaction with the environment and this effect should reduce the pumped charge. 
To demonstrate this effect, we have chosen an asymmetric pumping loop in the flux and gate voltage parameters and performed a numerical integration of Eqs. (\ref{eq:Eq1}), (\ref{eq:Eq2}), (\ref{eq:Eq1_sec}), and (\ref{eq:Eq2_sec}).
The results with and without  the secular approximation are presented in Fig. \ref{fig:Charge_Conservation}.
The pumped charge predicted with the secular approximation decreases when the coupling increases whereas if we include the non-secular terms, we obtain a result which is essentially immune to $g$.
Importantly, if the non-secular terms are included in the master equation, the two charges through the left and right junctions are equal but if we apply the secular approximation, we obtain a significant difference in these charges.
The difference leads to charge non-conservation since no charge can be accumulated on the island for the quasi-stationary solution.

We can also estimate analytically this charge non-conservation.
Let us define the charge asymmetry between the junctions as $\Delta Q=Q_L^{G,{\rm sec}}-Q_R^{G,{\rm sec}}$,
where $Q_k^{G,{\rm sec}}$ is the charge through junction $k$ described by Eq. (\ref{eq:secular-stationary-limit-zeroT}).
Using Eqs.~(\ref{eq:parameters}),~(\ref{eq:charge_eq}), and $J_R =- J_L  \sin \left(\gamma +\frac{\varphi }{2}\right)/\sin \left(\gamma -\frac{\varphi }{2}\right)$,
we obtain
\begin{eqnarray}
 \Delta Q  &=& \frac{2 e }{ \hbar} \int_0^{T_p}dt  \left(\frac{\hbar  \dot{\eta}}{2}-2 E_{12} \Im \mbox{m}(\delta \rho ) \right) .
 \label{eq:DeltaQ}
\end{eqnarray}
The first term in the integral gives  no contribution since $\eta(0)=\eta(T_p)$. 
This is itself an important result since it means that, if we keep the non-secular terms (i.e. $\delta \rho =0 $), we obtain 
immediately that the charge is conserved.

With the help of Eq. (\ref{eq:rho_sec}), the second term can be explicitly written as
\begin{eqnarray}
  \Im \mbox{m}(\delta \rho ) &=& \left( \frac{\Re \mbox{e}\left(w_{ge}\right) \Gamma }{2 \omega_0^2}+ \frac{\Im \mbox{m}\left(w_{ge}\right) 
 	\Gamma ^2}{4 \omega_0^3}\right).
\end{eqnarray}

Using Eqs.~(\ref{eq:parameters}), (\ref{eq:W}), and the definition of $\Gamma$ and $S(\omega)$ at zero-temperature given in Sec. ~\ref{sec:quasi-stationary} and~\ref{sec:gate_charge_noise}, respectively, we can write explicitly the  charge asymmetry as a function of time-dependent parameters
\begin{widetext}
\begin{eqnarray}
	 \Delta Q  &=& 2 e  g^2 \hbar \int_0^{T_p}dt \Huge[ 
	 -\dot{\gamma}  \left(\frac{1}{2} R \left(1-\eta^2\right)^2 +\frac{\left(1-\eta^2\right)^{3/2} S(0) \eta ^2}{2 E_{12}}\right) \nonumber \\
   && + \dot{\eta} g^2 \hbar \left(\frac{1}{2} R^2 \left(1-\eta ^2\right)^2-\frac{\left(\eta ^2-1\right) S^2(0)
   \eta ^4}{2 E_{12}^2}+\frac{R \left(1-\eta ^2\right)^{3/2} S(0) \eta^2}{E_{12}} \right) \Huge] .
	 \label{eq:ExplicitDeltaQ}
\end{eqnarray}
\end{widetext}

In this expression, we can address five contributions.
Since the third term depends only on $\eta$ and its derivative, it yields no contribution when integrated along the cycle.
The remaining terms can be calculated if we fix the phase to $\varphi=\pi/2$ and in the regime in which
$J_{i}^m \ll J_i^M\ll E_C$ and $\delta n_g^M, |\delta n_g^m| \gg J_i^M/ E_C$.
Using Eqs.~(\ref{eq:parameters}) and~(\ref{eq:W}), we can write Eq.~(\ref{eq:ExplicitDeltaQ}) in terms of the experimentally controlled parameters $J_L$, $J_R$, and $n_g$, and take advantage of the control cycle in Figure~\ref{fig:scheme}(b).
To have an idea of how the final terms look like we write the first term in Eq.~(\ref{eq:ExplicitDeltaQ}) as
\begin{equation}
A_1= \int_0^{T_{p}} dt \left[\frac{\left(\dot{J}_R J_L-\dot{J}_L J_R\right) \eta^2 (1-\eta^2)^\frac{3}{2}
g^2 R S_0}{ 2\left(J_L^2+J_R^2\right)^\frac{3}{2}}\right].
\end{equation}
It is convenient to divide the integration in six parts [see Fig.~\ref{fig:scheme} (b)]: for example, in the first path only $J_L$ depends on time.
The remaining $\eta$ function can be approximated, since,  for $J_{i}^m \ll J_i^M\ll E_C$, it tends to a box function switching between $\pm1$ [see Eq. (\ref{eq:parameters})].

\begin{figure*}
	\centering
		\includegraphics[scale=0.7]{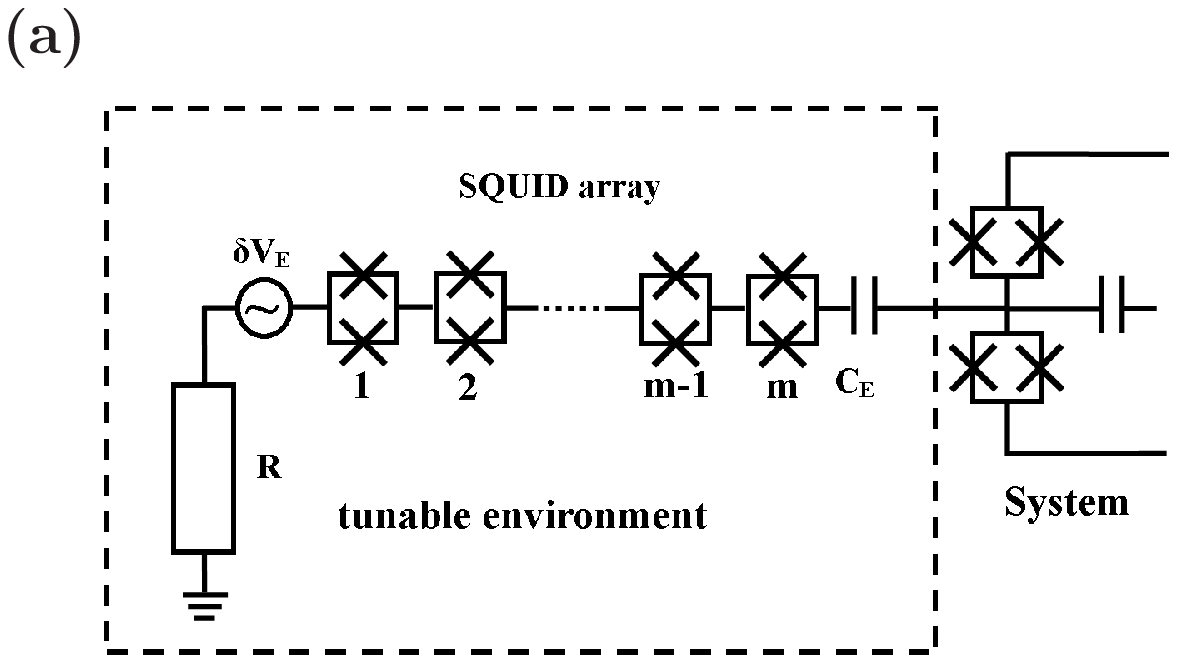}
		\includegraphics[scale=0.7]{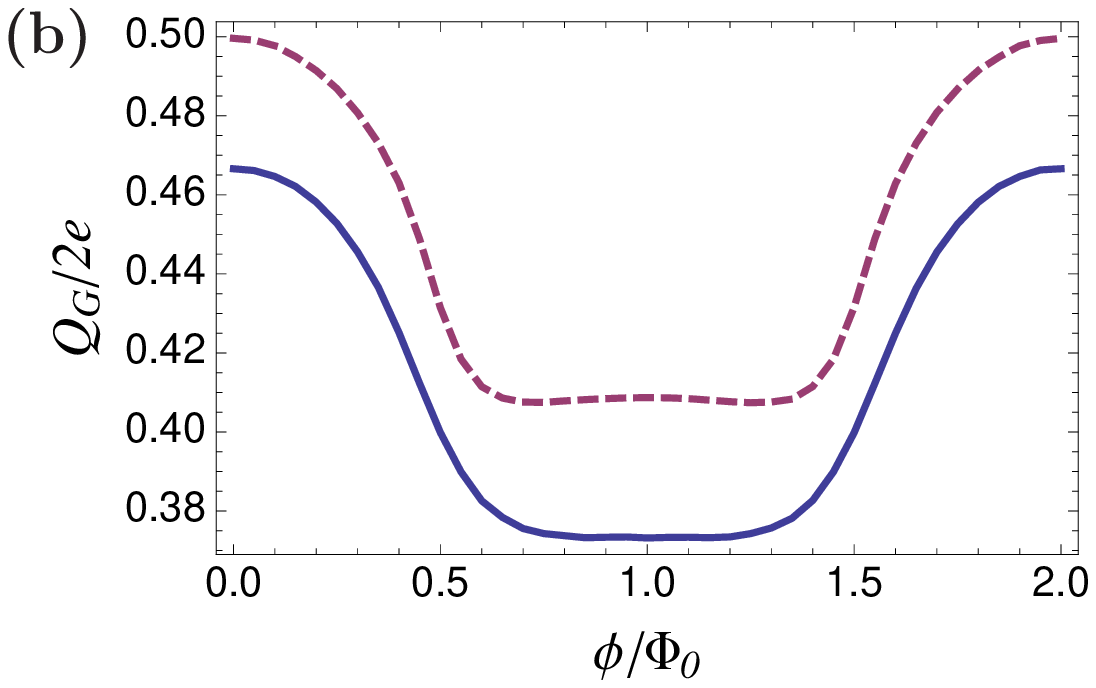}
	\caption{(Color online) (a) Engineered environment connected to the sluice pump.	
  The system-environment interaction is controlled by the flux through the SQUID array.
	(b) Pumped charge as a function of the flux through the SQUID array for
	$R_S=500~ \Omega$, $C_S=0.3$~fF (solid line) and
	 $R_S=1~{\rm k}\Omega$, $C_S=0.1$~fF (dashed line).
	The other parameters  are $f=10$ MHz, $\varphi=\pi/2$, $g=0.025$,  $m=100$ (number of SQUIDs in the array),  $C_E= 1$~fF,	$I_{C}=4$~nA, $R=1.5~{\rm k}\Omega$. In both cases, the simulations are performed with zero-temperature environment.
  }
	\label{fig:flux_noise}
\end{figure*}

With this further approximation, we obtain four leading contributions
\begin{eqnarray}
A_1&=&\frac{[J_L^m J_R^M+ J_R^m J_L^M ]  \left[(\delta n_g^M)^3-|\delta n_g^m|^3\right] g^2 \hbar ^2 S_0}{16 (E_C \delta n_g^m \delta n_g^M)^3}\nonumber\\ 
A_2&=&\frac{\left[J_R^m (J_L^M)^3+J_L^m (J_R^M)^3\right]  \left[(\delta n_g^M)^4-(\delta n_g^m)^4\right] g^2 R \hbar^2 }
{96 (E_C \delta n_g^m \delta n_g^M)^4} \nonumber\\ 
A_4&=& \frac{\left[(J_L^M)^2-(J_R^M)^2\right]  \left[(\delta n_g^m)^4+(\delta n_g^M)^4\right] g^4 \hbar ^3 S_0^2}{64 (E_C \delta n_g^m \delta n_g^M)^4} \nonumber\\
A_5&=& \frac{\left[(J_L^M)^4-(J_R^M)^4\right]  \left[(\delta n_g^m)^5+|\delta n_g^M|^5\right] g^4 \hbar ^3 S_0^2}{128 (E_C \delta n_g^m \delta n_g^M)^5},
\label{eq:As}
\end{eqnarray}
where $\Delta Q= \sum_{i=1,5} A_i$ with $A_3=0$.
The first two contributions are present only if the loop is asymmetric in the $\delta n_g$ parameter (i.e. $\delta n_g^M \neq |\delta n_g^m|$) whereas $A_4$  and $A_5$ yield non-vanishing  contributions if the loop is asymmetric in the flux parameters (i.e. $J_L^M \neq J_R^M$). \cite{JNote}
If the loop is asymmetric in both parameters as the one used in the simulations of Fig.~\ref{fig:Charge_Conservation}, all the $A_i$ terms in Eq.~(\ref{eq:As}) contribute to $\Delta Q$.

\subsection{Environment engineering}
\label{sec:engineered_env}

The experimental set-up described in this paper has already been  used for the measurement of the Berry phase in adiabatic pumping of Cooper pairs. \cite{mottonen06,mottonen08}
To study the effects of the environment in the pumping, it is beneficial to control the coupling between the system and its environment.
This can be achieved by adding to the already present intrinsic environment an experimentally controlled artificial noise source.
Similar approaches have been developed in different physical systems for both testing the robustness of the studied quantum system and the cross-over between the quantum and classical behavior. \cite{myatt00, kielpinki01}
Here, we present a way to engineer {\it in situ} the environment noise in the superconducting circuits discussed above.

The environment is controlled by the circuit in Fig.\ref{fig:scheme} which is composed of a thermal resistor $R$ with noise voltage spectrum $\delta V_E$.
We place an array of  $m$ SQUIDs between the resistor and the island of the sluice as shown in Fig. \ref{fig:flux_noise}.
The array can be represented as a series of $RLC$ parallel circuits with resistance $R_S$, inductance $L$, and capacitance $C_S$.
The impedance of a single SQUID is $Z_{RLC}= \frac{R_S Z_{C_S} Z_L}{R_S Z_{C_S}+Z_L Z_{C_S}+R_S Z_L}= \frac{i L(\phi) \omega  R_S}{i L(\phi) \omega +R_S \left(1-L(\phi) \omega ^2   C_S \right)}$, where $L(\phi)= L_0/\cos (\pi  \phi/\Phi_0 )$,  $Z_L$ and $Z_{C_S}$ are the impedances associated to the inductance $L$ and capacitance $C_S$, respectively, $\phi/\Phi_0$ is the flux through the SQUID in units of the flux quantum, and $L_0=\hbar/(2 \pi e I_{C} )$ where $I_{C}$ is the maximum critical current of the SQUID.
Note that the Josephson inductance $L(\phi)$ can be modulated by applying flux through the SQUID.

The total series impedance of the circuit with respect to the noise source in Fig. \ref{fig:flux_noise} is $Z_{\rm tot}=R+Z_{C_E}+m Z_{RLC}$, and the voltage at the capacitor $C_E$ is $\tilde{V}=V Z_{C_E}/Z_{\rm tot}$.
Here, we neglect the impedance of the sluice and possible parasitic capacitance directly to ground inside the tunable environment.
In this way, the gate voltage noise spectrum perceived by the system is 
\begin{widetext}
\begin{equation}
  \tilde{S}(\omega,\phi) = \left |\frac{Z_{C_E}}{Z_{\rm tot}} \right| ^2 S(\omega) = 
  \frac{{S(\omega)}}{
   \left(1-m \frac{L(\phi)  \omega ^2
   R_S^2 C_E-m L^2(\phi) \omega ^4 R_S^2
   C_S C_E}{L^2(\phi) \omega ^2+R_S^2 \left(1-L(\phi)
   \omega ^2 C_S \right){}^2}\right)^2+
   \left( \frac{m L^2(\phi)  R_S C_E \omega ^3}{L^2(\phi) \omega
   ^2+R_S^2 \left(1-L(\phi) \omega ^2 C_S\right){}^2}+ R
   C_E \omega \right)^2}.
 \label{eq:effective_spectrum}
\end{equation}
\end{widetext}
Controlling the flux through the SQUID array, we can change the spectrum $\tilde{S}(\omega,\phi)$ and hence we
effectively tune the coupling between the system and the environment.

The pumped charge expected as a function of the flux through the SQUIDs in the array is presented in Fig.~\ref{fig:flux_noise}.
The results are obtained solving the master Eqs.~(\ref{eq:Eq1}) and~(\ref{eq:Eq2}) numerically with the effective spectrum given by Eq.~(\ref{eq:effective_spectrum}).
This modulation of the the pumped charge could be large enough to be experimentally observable.

\section{Conclusions}
\label{sec:conclusions}

In conclusion, we have presented a detailed study of the effect of environmental noise on the adiabatic evolution of a two-level quantum system.
In the zero-temperature limit, we find that the adiabatic evolution in the ground state is not influenced by the presence  of the noise implying that the evolution is robust against environmental noise.

These results are obtained taking consistently into account the contributions in the master equation for the dynamics of the density matrix.
In particular, we keep the non-secular terms and the terms describing the combined effect of the environment and the drive.
This step is important to obtain the robustness of the ground state dynamics and, at the same time, to guarantee charge conservation.

We have tested our theory by applying it to Cooper pair pumping in presence of system-environment coupling.
The numerical simulations confirm the theoretical prediction about the robustness of the ground-state pumping and suggest that in the non-adiabatic regime, relaxation can help to restore the desired ground-state evolution.
In the same system, we studied the pumped charge in the secular approximation and observed that the pumped charge decreases with increasing system-environment coupling strength.
Furthermore, the pumped charges through the first and the second junction become different implying unphysical charge non-conservation.
Thus we conclude that the secular approximation in this form cannot be pursued in adiabatic evolution.

We have proposed a way to engineer the environmental noise in the discussed physical system.
This allows us to modify the effective spectrum of the environment by changing the magnetic flux through an array of SQUIDs.
The consequent change in the pumped charge could be experimentally observable.

The results obtained concerning the robustness of adiabatic ground-state evolution are quite general and  it would be interesting to extend our analysis.
For example, adiabatic quantum computing \cite{farhi01} is based on the adiabatic evolution in the ground state.
Another possible extension is the construction of decoherence-free logical gates in the geometric quantum computation paradigm. \cite{zanardi99}
In this case, the evolution occurs in a degenerate subspace and the manipulation of the quantum state is performed with the help of adiabatic evolution.
In many of the proposals for geometric quantum gates, one of the main disadvantages is that the degenerate subspace is not the one with the lowest-energy. This likely  leads to problems in controlling the relaxation to the effective ground state.
However, geometric quantum computing can, in principle, be performed in the ground-state manifold. \cite{pirkkalainen10}
A detailed study of its robustness can be carried out by extending the theory presented here beyond the two-level approximation. However, this is left for future research.
\\
\\
\appendix
\section{Derivation of the master equation}

The master equation~(\ref{me1}) is obtained with a development taking into account all the terms up to the order $w V V$.
As in standard derivation of the master equation \cite{cohen-tannoudji}, we assume that the density matrix of the environment is stationary and that the average of $V$ over the environment degrees of freedom vanishes.

Denoting the total density matrix of the system and  environment as $\tilde{\rho}^{\rm tot}(t)$ and employing the transformation to the adiabatic basis as $\tilde{\rho}_I^{\rm tot}=D^\dagger \tilde{\rho}^{\rm tot} D$, the von Neumann equation in the interaction picture reads
\begin{equation} \label{eq1}
\dot{\tilde{\rho}}_I^{\rm tot}(t)=\frac{i}{\hbar}[\tilde{\rho}_I^{\rm tot}(t), \hbar w_I(t)+\tilde{V}_I(t)].
\end{equation}
Notice that, in the weak coupling and in the adiabatic limit, $w_I(t)$ and $\tilde{V}_I(t)$ are perturbative contributions of different order.
Tracing over the degree of freedom of the environment Eq.~(\ref{eq1}) becomes
\begin{equation}
\dot{\tilde{\rho}}_I(t)=i [\tilde{\rho}_I(t), w_I(t)]+ \frac{i}{\hbar} {\rm Tr}_E \big{\{}[\tilde{\rho}_I^{\rm tot}(t),\tilde{V}_I(t)]\big{\}},
\label{eq:eq2}
\end{equation}
where $\tilde{\rho}_I^{\rm tot}= {\rm Tr}_E \big{\{} \tilde{\rho}_I^{\rm tot} \big{\}}$.

Together with Eq.~(\ref{eq1}), we employ the identity
\begin{equation}
\tilde{\rho}_I^{\rm tot}(\tau)=\tilde{\rho}_I^{\rm tot}(\tau_1)+\int_{\tau_1}^\tau d\tau' \dot{\tilde{\rho}}_I^{\rm tot}(\tau').
\label{eq:formal_rho}
\end{equation} 
Using iteratively Eqs.~(\ref{eq1}) and (\ref{eq:formal_rho}) we can obtain a perturbation expansion of Eq.~(\ref{eq:eq2}).

Substituting Eq.~(\ref{eq:formal_rho}) in the last term in Eq.~(\ref{eq:eq2}) we have
\begin{eqnarray}
\dot{\tilde{\rho}}_I(t)&=&i[\tilde{\rho}_I(t), w_I(t)]
 +\frac{i}{\hbar} {\rm Tr}_E \big{\{}[\tilde{\rho}_I^{\rm tot}(0),\tilde{V}_I(t)] \nonumber \\
&& +\int_0^t dt'[\dot{\tilde{\rho}}_I^{\rm tot}(t'),\tilde{V}_I(t)]\big{\}}.
\end{eqnarray}

Since the average of $V$ over the environment degrees of freedom vanishes,  ${\rm Tr}_E \big{\{}[\tilde{\rho}_I^{\rm tot}(0),\tilde{V}_I(t)] \big{\}}=0$ and using Eq.~(\ref{eq1}), we obtain
\begin{widetext}
\begin{eqnarray}
\dot{\tilde{\rho}}_I(t)&=&i [\tilde{\rho}_I(t), w_I(t)]-\frac{1}{\hbar^2} {\rm Tr}_E
\big{\{}\int_0^t dt'\big[ [\tilde{\rho}_I^{\rm tot} (t'),\hbar w_I(t')],\tilde{V}_I(t)\big]\big{\}} 
-\frac{1}{\hbar^2} {\rm Tr}_E\big{\{}\int_0^t dt'\big[ [\tilde{\rho}_I^{\rm tot} (t'),\tilde{V}_I(t')],\tilde{V}_I(t)\big]\big{\}}.
\end{eqnarray}
\end{widetext}
The term $\tilde{\rho}_I^{\rm tot} (t')$ can be transformed using Eq. (\ref{eq:formal_rho}); in particular, we substitute 
$\tilde{\rho}_I^{\rm tot} (t')=\tilde{\rho}_I^{\rm tot} (0)+\int_{0}^{t'} dt^{\prime \prime} \dot{\tilde{\rho}}_I^{\rm tot} (t^{\prime \prime} )$  and $\tilde{\rho}_I^{\rm tot} (t')=\tilde{\rho}_I^{\rm tot} (t)-\int_{t'}^{t} dt^{\prime \prime} 
\dot{\tilde{\rho}}_I^{\rm tot} (t^{\prime \prime} )$ in the second and in the third term on the right, respectively.
Consistently, we have that ${\rm Tr}_E \{ \big[[\tilde{\rho}_I^{\rm tot}(0),w_I(t')],\tilde{V}_I(t)\big]\} =0$ and hence
\begin{widetext}
\begin{eqnarray}
\dot{\tilde{\rho}}_I(t)&=&i  [\tilde{\rho}_I(t), w_I(t)] 
-\frac{1}{\hbar^2} {\rm Tr}_E\big{\{} \int_0^t dt'\big[[\tilde{\rho}_I^{\rm tot}(t),\tilde{V}_I(t')],\tilde{V}_I(t)\big]\big{\}}
-\frac{1}{\hbar^2} {\rm Tr}_E\big{\{} \int_0^t dt' \int_0^{t'}dt'' \big[[ \dot{\tilde{\rho}}_I^{\rm tot} (t''), \hbar w_I(t')],\tilde{V}_I(t)\big]\big{\}}\nonumber \\
&&+\frac{1}{\hbar^2} {\rm Tr}_E\big{\{}  \int_0^t dt' \int_{t'}^{t} dt'' \big[[ \dot{\tilde{\rho}}_I^{\rm tot}(t''),\tilde{V}_I(t')],\tilde{V}_I(t)\big]\big{\}}.
\end{eqnarray}
\end{widetext}

Using Eq. (\ref{eq1}) we eliminate $\dot{\tilde{\rho}}_I^{\rm tot}(t'')$ from the above equation and, keeping the terms up to order $w V V$, we obtain 
\begin{widetext}
\begin{eqnarray}
\dot{\tilde{\rho}}_I(t)&=&i[\tilde{\rho}_I(t), w_I(t)] 
-\frac{1}{\hbar^2} {\rm Tr}_E\big{\{} \int_0^t dt'\big[[\tilde{\rho}_I^{\rm tot}(t),\tilde{V}_I(t')],\tilde{V}_I(t)\big]\big{\}}
-\frac{i}{\hbar^2} {\rm Tr}_E\big{\{} \int_0^t dt' \int_0^{t'}dt'' \Big[ \big[ [\tilde{\rho}_I^{\rm tot} (t''),\tilde{V}_I(t'')], w_I(t')\big],\tilde{V}_I(t)\Big]\big{\}}\nonumber \\
&&+\frac{i}{\hbar^2} {\rm Tr}_E\big{\{}  \int_0^t dt' \int_{t'}^t dt'' \Big[ \big[ [ \tilde{\rho}_I^{\rm tot}(t''),  w_I(t'') ],\tilde{V}_I(t') \big],\tilde{V}_I(t)\Big]\big{\}}.
\label{eq:last_eq}
\end{eqnarray}
\end{widetext}

The third and fourth terms on the right are both of order $w V V$, namely, the highest order in our expansion.
The last step in our derivation is to use Eq.~(\ref{eq:formal_rho}) to substitute $\tilde{\rho}_I^{\rm tot}(t'')$ with $\tilde{\rho}_I^{\rm tot} (t)+\int_{t}^{t''} dt''' 
\dot{\tilde{\rho}}_I^{\rm tot} (t''')$; however, since the terms with derivative of $\tilde{\rho}_I^{\rm tot}$ give contributions either of order $w$ or $V$, they can be neglected. 
Thus, we can effectively substituted $\tilde{\rho}_I^{\rm tot}(t'')$ with  $\tilde{\rho}_I^{\rm tot}(t) $ in Eq.~(\ref{eq:last_eq}) without introducing further approximations. 
The master equation~(\ref{me1}) is obtained by rearranging the integration limits and the commutators of the last two terms of the resulting equation.

The authors thank  J. Ankerhold, V. Gramich,  A. Shnirman, V. Brosco and I. Kamleitner  for insightful discussions. We have received funding from the European Community's Seventh Framework Programme under Grant Agreement No. 238345 (GEOMDISS). We acknowledge Academy of Finland and Emil Aaltonen Foundation for financial support.


\begin{thebibliography}{99}
\bibitem{farhi01} E. Farhi,  J. Goldstone, S. Gutmann, J. Lapan, A. Lundgren,  D.Preda, Science {\bf 292}, 472 (2001).
\bibitem{ad_vs_dyn} D. Aharonov, W. van Dam, J. Kempe, Z. Landau, S. Lloyd, O. Regev  SIAM J. Comput.  {\bf 37}, 166 (2007). A.Mizel, D. A. Lidar and M. Mitchell, Phys. Rev. Lett. {\bf 99}, 070502 (2007).
\bibitem{berry84} M. V. Berry, Proc. R. Soc. A {\bf 392}, 45 (1984).
\bibitem{wilczek84} F. Wilczek and A. Zee, Phys. Rev.  Lett. {\bf 52}, 2111 (1984).  
\bibitem{falci00}  G. Falci, R. Fazio, G. M. Palma, J. Siewert, and V. Vedral, Nature {\bf 407}, 355 (2000).
\bibitem{zanardi99}  P. Zanardi and M. Rasetti, Phys. Lett. A {\bf 264}, 94 (1999).
\bibitem{unanyan99} R. G. Unanyan, B. W. Shore, and K. Bergmann, Phys. Rev. A {\bf 59}, 2910 (1999).
\bibitem{duan01} L.-M. Duan, J. I. Cirac and P. Zoller, Science {\bf 292}, 1695, (2001).
\bibitem{dechiara03} G. De Chiara and G. M. Palma, Phys. Rev. Lett. {\bf 91}, 090404 (2003).  
\bibitem{solinas04} P. Solinas, P. Zanardi, and N. Zangh\'i, Phys. Rev. A {\bf 70}, 042316 (2004).


\bibitem{parodi07} D. Parodi, M. Sassetti, P. Solinas, P. Zanardi, and N. Zangh\'i, Phys. Rev. A {\bf 73}, 052304 (2006).
D. Parodi, M. Sassetti, P. Solinas, and N. Zangh\'i, Phys. Rev. A {\bf 76}, 012337 (2007).
\bibitem{florio06} G. Florio, P. Facchi, R. Fazio, V. Giovannetti, S. Pascazio,  Phys. Rev. A {\bf 73}, 022327 (2006).

\bibitem{cohen-tannoudji} C. Cohen-Tannoudji, J. Dupont-Roc, and G. Grynberg, {\it Atom-Photon Interactions} (Wiley, New York 1992).
\bibitem{pekola09} J. P. Pekola, V. Brosco, M. M\"ott\"onen, P. Solinas, and A. Shnirman, Phys. Rev. Lett. {\bf 105}, 030401 (2010).

\bibitem{calarco03} T. Calarco, A. Datta, P. Fedichev, E. Pazy, and P. Zoller,  Phys. Rev. A {\bf 68} , 012310 (2003).
\bibitem{roszak05} K. Roszak, A. Grodecka, P. Machnikowski, and T. Kuhn, Phys. Rev. B {\bf 71} , 195333 (2005).
\bibitem{caillet07} X. Caillet and C. Simon, Eur.Phys.J.D 42, 341 (2007).

\bibitem{pirkkalainen10} J.-M. Pirkkalainen, P. Solinas, J. P. Pekola, and M. M\"ott\"onen, Phys. Rev. B {\bf 81}, 174506 (2010) 



\bibitem{niskanen03} A. O. Niskanen, J. P. Pekola, and H. Sepp\"a, Phys. Rev. Lett. {\bf 91}, 177003 (2003).
\bibitem{niskanen05} A. O. Niskanen, J. M. Kivioja, H. Sepp\"a, and J. P. Pekola, Phys. Rev. B. {\bf 71}, 012513 (2005).


\bibitem{sarandy05} M. S. Sarandy and D. A. Lidar, Phys. Rev. Lett. {\bf 95}, 250503 (2005).
\bibitem{whitney05}  R. S. Whitney, Y. Makhlin, A. Shnirman, and Y. Gefen, Phys. Rev. Lett. {\bf 94}, 070407 (2005).
\bibitem{note_m} Notice that, to write Eqs. (\ref{eq:Eq1}) and (\ref{eq:Eq2}) in a more compact form,  we have included $\hbar$ in the definition of $m_1$ and $m_2$.

\bibitem{lamb_shift_note} J. Ankerhold and V. Gramich private communication.


\bibitem{aunola03} M. Aunola and J. J. Toppari, Phys. Rev. B {\bf 68}, 020502(R) (2003).
\bibitem{mottonen08} M. M\"ott\"onen, J. J. Vartiainen, and J. P. Pekola, Phys. Rev. Lett. {\bf 100}, 177201 (2008).
\bibitem{mottonen06}M. M\"ott\"onen, J. P. Pekola, J. J. Vartiainen, V. Brosco, and F. W. J. Hekking, Phys. Rev. B {\bf 73}, 214523 (2006).
\bibitem{leone08}R. Leone, L. P. Levy, and P. Lafarge, Phys. Rev. Lett. {\bf 100}, 117001 (2008).


\bibitem{salmilehto10} J. Salmilehto. P. Solinas, M. M\"ott\"onen, Phys. Rev. A {\bf 82}, 062112 (2010).


\bibitem{vartiainen07}J. J. Vartiainen, M. M\"ott\"onen,  and J. P. Pekola, Appl. Phys. Lett. {\bf 90}, 082102 (2007).
\bibitem{makhlin01} Y. Makhlin, G. Sch\"on, and A. Shnirman, Rev. Mod. Phys. {\bf 73}, 357 (2001).

\bibitem{JNote} We recall that these are the leading contributions when $J_i^M \gg J_i^m$ and $J_L^M \neq J_R^M$.

\bibitem{myatt00} C. J. Myatt, B. E. King, Q. A. Turchette, C. A. Sackett, D. Kielpinski, W. M. Itano, C. Monroe, and D. J. Wineland, Nature {\bf 403}, 269 (2000). 
\bibitem{kielpinki01} D. Kielpinski,  V. Meyer,  M. A. Rowe,  C. A. Sackett,  W. M. Itano,  C. Monroe,  D. J. Wineland, Science, {\bf 291}, 1013 (2001).

\end{thebibliography}
\end{document}